\documentclass[lettersize,journal]{IEEEtran}
\usepackage{amsmath,amsfonts}
\usepackage{algorithmic}
\usepackage{algorithm}
\usepackage{array}
\usepackage[caption=false,font=normalsize,labelfont=sf,textfont=sf]{subfig}
\usepackage{textcomp}
\usepackage{stfloats}
\usepackage{url}
\usepackage{verbatim}
\usepackage{graphicx}
\usepackage{cite}
\usepackage{xcolor}
\usepackage{soul}
\usepackage{ulem}

\begin{document}

\title{Practical Approaches to Quantifying Intra-Pair Skew Impact via Insertion Loss Deviation}

\author{\IEEEauthorblockN{David Nozadze\IEEEauthorrefmark{1}\IEEEauthorrefmark{2},
		Zurab Kiguradze\IEEEauthorrefmark{1},
		Amendra Koul\IEEEauthorrefmark{1},
		Sayed Ashraf Mamun\IEEEauthorrefmark{1}, and
		Mike Sapozhnikov\IEEEauthorrefmark{1}}
		
	\IEEEauthorblockA{\IEEEauthorrefmark{1} Cisco Systems Inc., San Jose, CA, USA }
	
	\IEEEauthorblockA{\IEEEauthorrefmark{2} dnozadze@cisco.com}}

        % <-this % stops a space
%\thanks{This paper was produced by the IEEE Publication Technology Group. They are in Piscataway, NJ.}% <-this % stops a space
%\thanks{Manuscript received April 19, 2021; revised August 16, 2021.}}

\markboth{IEEE TRANSACTIONS ON SIGNAL AND POWER INTEGRITY, VOL. num, 2025}%
{Shell \MakeLowercase{\textit{et al.}}: A Sample Article Using IEEEtran.cls for IEEE Journals}


\maketitle

\begin{abstract}
The surge in AI workloads and escalating data center requirements have created demand for ultra-high-speed interconnects exceeding 200 Gb/s. As unit intervals (UI) shrink, even a few picoseconds of intra-pair skew can significantly degrade serializer-deserializer (SerDes) performance. To quantify the impact of intra-pair skew, conventional time-domain methods are often unreliable for coupled interconnects due to skew variations across voltage levels, while frequency-domain approaches frequently fail to address reciprocity and symmetry issues. This can result in channels that meet skew specifications in one direction but not the other, despite the inherently reciprocal nature of skew impact. To address these limitations, we introduce two new reciprocal parameters for quantifying intra-pair skew effects: Skew-Induced Insertion Loss Deviation (SILD) and its complementary Figure of Merit (\text{FOM\_SILD}). Measurements conducted using 224 Gb/s SerDes IP and a variety of channels with different intra-pair skews demonstrate a strong correlation between \text{FOM\_SILD} and bit error rate (BER). Results show that when \text{FOM\_SILD} is below 0.2-0.3 dB, BER remains stable, indicating minimal signal integrity degradation; however, BER increases noticeably as \text{FOM\_SILD} exceeds 0.3 dB. Statistical analysis across more than 3,000 high-speed twinax cables reveals that the majority exhibit \text{FOM\_SILD} values less than ~0.1 dB, underscoring the practical relevance of the proposed metrics for high-speed interconnect assessment.
\end{abstract}


\section{Introduction}
\begin{figure}[!t]
	%\centering
	\includegraphics[width=3.5in]{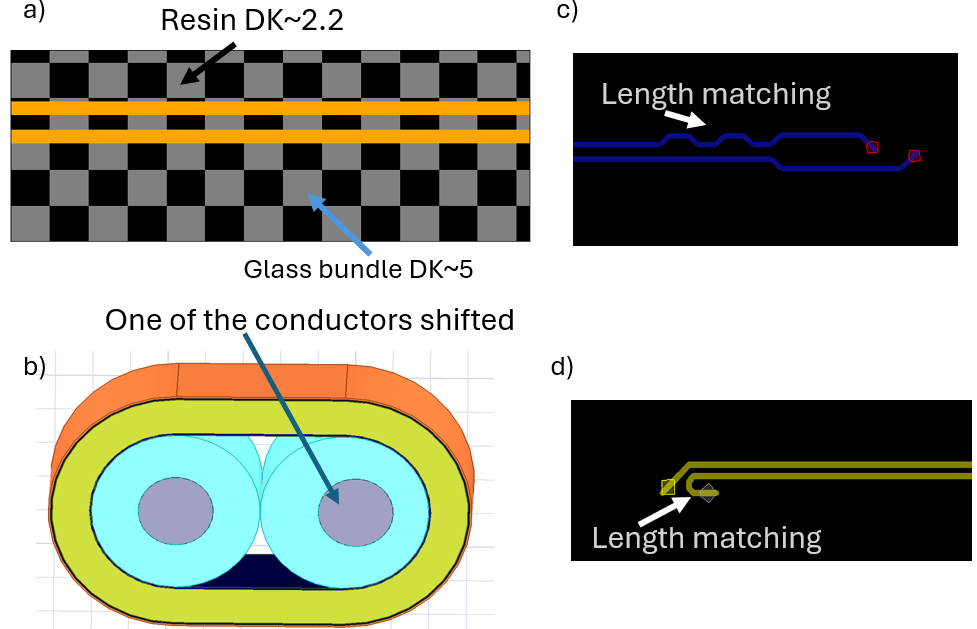}
	% where an .eps filename suffix will be assumed under latex, 
	% and a .pdf suffix will be assumed for pdflatex; or what has been declared
	% via \DeclareGraphicsExtensions.
	\caption{a) Differential traces with respect to the glass fiber weave on the PCB.
		b) 3D drawing of a high-speed twinax cable with one conductor shifted.
		c-d) Examples of two different types of physical length matching for differential traces on a PCB.}
	\label{fig_1}
	%\vspace{-1.8em}
\end{figure}

As data rates continue to rise, it becomes increasingly important to assess all factors within a signal channel that can affect signal integrity. At speeds above 200 Gbps, intra-pair differential skew emerges as a major performance-limiting factor for high-speed SerDes links. The intra-pair differential skew is the difference in arrival times between the two single-ended signals in a differential pair. This effect is typically attributed to asymmetry between the positive (P) and negative (N) traces of the differential pair. However, even when the physical lengths of the P and N traces are perfectly matched, skew can still result from other asymmetries present in the signal path. For example:
1) As shown in Fig.~\ref{fig_1}a, due to Printed Circuit Board (PCB) material non-homogeneity, the P line may encounter a different proportion of glass than the N line. Since the dielectric constants of glass and resin differ, the signals on the P and N traces propagate at different velocities, resulting in intra-pair skew.
2) In Fig.~\ref{fig_1}b, intra-pair skew can occur in twinax cables if the wire is off-center or due to other geometric imperfections.
3) As illustrated in Figs.~\ref{fig_1}c and 1d, non-symmetrical routing between the P and N lines can also lead to intra-pair skew.
While numerous studies have attempted to measure and quantify intra-pair skew, a comprehensive methodology for accurately predicting its impact on SerDes performance is still missing  \cite{2014_Tian_simp,2010_Miller_DC,2017_Nozadze_epeps,2007_Loyer_CT,2017_Baek_DC,Nalla_2017_EMC,Nozadze_2017_EMC,2021_Moon_spi, 2018_Koul_DC}.

To overcome the limitations of non-reciprocity and excitation dependence inherent in conventional metrics, this work proposes a mathematically robust and reciprocal framework for quantifying intra-pair skew impact. We introduce two new metrics: the Skew-Induced Insertion Loss Deviation (SILD) and its scalar Figure of Merit (\text{FOM\_SILD}). Unlike existing time-domain or frequency-domain methods, these metrics provide a consistent evaluation of the channel regardless of signal direction, directly relating skew-induced phase mismatches to the distortion of differential insertion loss—a parameter to which SerDes receivers are highly sensitive. The contribution of this work is three-fold: first, the derivation of a de-skewing algorithm for S-parameters; second, the experimental validation of these metrics through 224 Gb/s SerDes bit error rate (BER) measurements; and third, a large-scale statistical validation across 3,000+ industrial cable channels to establish practical qualification thresholds for ultra-high-speed interconnects.

The paper is organized as follows. In Sec. II, we review existing conventional methods and metrics used for intra-pair skew quantification and discuss their limitations. Our new metrics are introduced in Sec. III, along with their mathematical formulations. Sec. IV presents the correlation between \text{FOM\_SILD} and the measured BER trends of 224 Gbps SerDes IP. In Sec. V, we present the statistical distribution of \text{FOM\_SILD} across over 3,000 Direct Attach Copper (DAC) cable measurements. Conclusions are provided in Sec. VI.

\section{Conventional Metrics for intra-pair skew}
In this section, we review conventional methods and metrics used to quantify intra-pair skew impact on high-speed channels and highlight some of their critical limitations.

\subsection{Time Domain Analysis}\label{sec_TDT}

\begin{figure}[!t]
	%\centering
	\includegraphics[width=3.5in]{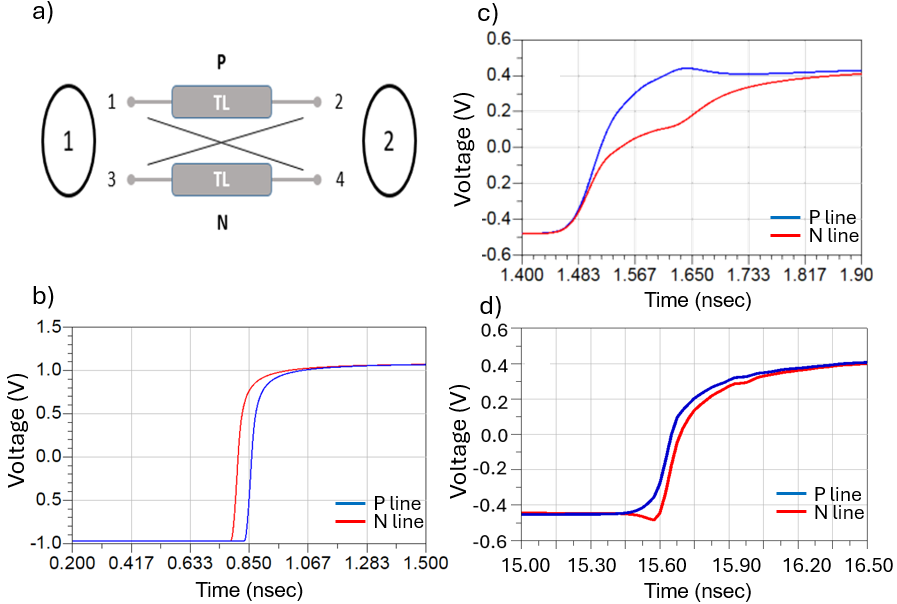}
	% where an .eps filename suffix will be assumed under latex, 
	% and a .pdf suffix will be assumed for pdflatex; or what has been declared
	% via \DeclareGraphicsExtensions.
	\caption{a) Schematic of Coupled Transmission Line: Ports labeled as 1 and 2 represent mixed-mode ports, while ports labeled as 1 through 4 correspond to single-ended ports. The P/N lines form a differential pair.		
		 The time-domain transmission response to a step excitation through P/N lines is shown for: b) a PCB stripline, c) a PCB microstrip line, and d) twinax cables.}
	\label{fig_2}
	%\vspace{-1.8em}
\end{figure}

A conventional approach to quantifying intra-pair skew in differential transmission lines is through the time-domain transmission (TDT) response to a step excitation applied to the P and N conductors (Fig.~\ref{fig_2}a). The skew is then defined as the relative displacement between their rising edges. This method is effective for loosely coupled transmission lines, such as PCB striplines, where the rising edges of the P and N signals remain parallel (Fig.~\ref{fig_2}b). In this case, the skew corresponds to a simple time-delay skew, since even- and odd- modes propagate with identical velocities. Consequently, the time shift between edges, measured across different voltage thresholds, remains constant, providing a well-defined measure of skew.

The single-ended insertion losses for the P and N lines can be written as:
\begin{align*}
	s_{12} &= \exp(-\gamma L_{12}) \cdot \exp(-2i\pi f t_{12}) \\
	s_{21} &= \exp(-\gamma L_{21}) \cdot \exp(-2i\pi f t_{21})
\end{align*}
where:
\begin{itemize}
	\item $\gamma$ is the complex propagation constant of single P/N lines.
	\item $L_{12}$ and $L_{21}$ are lengths of the P and N lines, respectively. Assume that the P and N lines have the same attenuation
	$\exp(-\gamma L_{12})\approx \exp(-\gamma L_{21})$.
	\item $f$ is the frequency.
	\item $t_{12}$ and $t_{21}$ are the time delays associated with the P and N lines, respectively.
\end{itemize}
For simplicity, assuming that forward coupling is negligible compared to the through S-parameters. The differential insertion loss can be derived from $s_{12}$ and $s_{21}$ as follows:
\begin{align*}
	|S_{dd21}| &\approx |s_{12} + s_{21}|/2 \approx \exp(-\gamma L_{12}) \left| \cos\left( \pi f t_{\text{skew}} \right) \right|,
\end{align*}
where $t_{\text{skew}} = |t_{12} - t_{21}|$ is the time skew between the P and N lines.
 
Therefore, the differential insertion loss due to skew $t_{\rm{skew}}$, as a function of frequency $f$, is
\begin{align}{\label{skewloss}}
	\text{SkewLoss}(f)=dB(cos(\pi f t_{\rm{skew}}))\,,
\end{align}
as reported in the \cite{Nozadze_2017_EMC}.

Note that this loss is not caused by dielectric dissipation or conductor loss. Instead, it occurs because asymmetry between the lines causes some energy to convert from differential mode to common mode, thereby reducing the differential signal.

For strongly coupled transmission lines, the situation is more complex. The rising edges of the P and N conductors are no longer parallel, and their displacement varies across the transition.
Rise-Time Skew: Fig.~\ref{fig_2}c illustrates the TDT response of coupled microstrip structures, where the relative delay increases toward the end of the transition. This phenomenon is referred to as rise-time skew, and it arises because the odd mode propagates faster than the even mode. Prior work \cite{2017_Nozadze_epeps} shows that in this case, the differential insertion loss does not attenuate as rapidly as in loosely coupled structures.
Amplitude Skew: Fig.~\ref{fig_2}d  shows the TDT response of dual-extruded twinax cables, where a large displacement is observed at the beginning of the transition. This case is referred to as amplitude skew, and here the even mode propagates faster than the odd mode \cite{2023_Liu_IEEE_TR}.

In real systems, a combination of time-delay, rise-time, and amplitude skews is typically present. As a result, the rising edges are not parallel, and their relative displacement varies depending on the voltage threshold. This variability makes a skew definition based solely on edge displacement ambiguous and impractical.

\begin{figure}[!t]
	%\centering
	\includegraphics[width=3.5in]{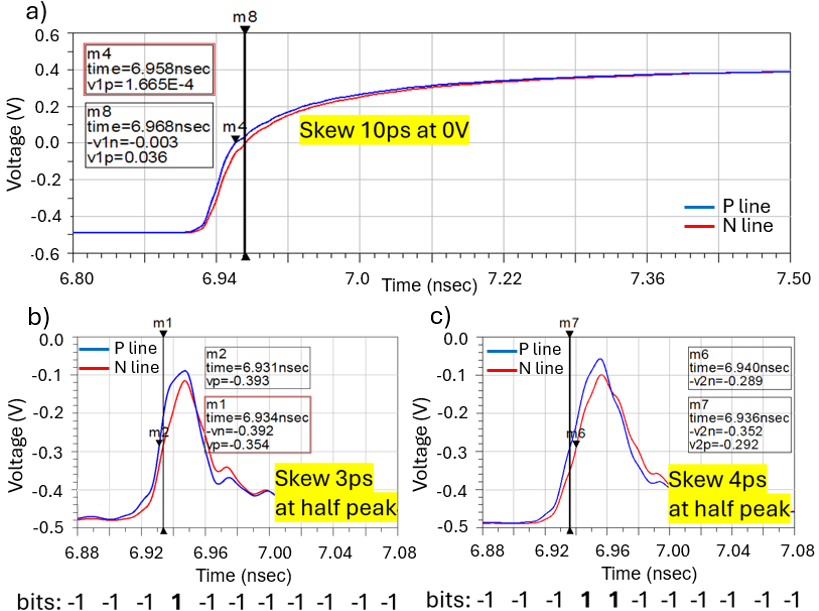}
	% where an .eps filename suffix will be assumed under latex, 
	% and a .pdf suffix will be assumed for pdflatex; or what has been declared
	% via \DeclareGraphicsExtensions.
	\caption{The measured time-domain transmission response of the 224 Gbps twinax cable through P/N lines is shown for (a) a step excitation, (b) single-bit pulses, and (c) two-bit pulses.}
	\label{fig_3}
	%\vspace{-1.8em}
\end{figure}

Fig.~\ref{fig_3} presents the TDT responses for a DAC channel that includes PCB fixtures, connectors, and twinax cables under different excitation:
(a) step input,
(b) single-bit pulse, and
(c) two-bit long pulse.
The observed skew magnitude differs across these excitations. This demonstrates that skew is not only voltage-level dependent but also strongly dependent on the excitation waveform.

The TDT-based approach provides a straightforward skew definition for loosely coupled structures, where even- and odd-modes propagate at the same velocity. However, for strongly coupled transmission lines, the presence of rise-time and amplitude skew, combined with excitation dependence, renders the TDT method unreliable for intra-pair skew quantification.

\subsection{Frequency Domain Skew}\label{sec_FDS}

In this section,  intra-pair skew is defined in frequency domain using Transmission Line (TL) theory. To do so, coupled TL shown in Fig.~\ref{fig_1} a) is considered. The intra-pair phase skew at differential port 2 is defined as

\begin{align}{\label{eq_skew1}}
	 t_{\rm{skew,21}}=t_{1,2}-t_{2,2}\,,
\end{align}
where
\[
%\begin{align}
t_{1,2}=\frac{{{phase}}(S_{sd21})}{2\pi f}  \hspace{5.5mm}\text{and}\hspace{5.5mm} t_{2,2}=\frac{{{phase}}(S_{sd41})}{2\pi f}\,,
\]
%\end{align} 
are time delays corresponding to the propagation of the signal from mixed-mode port 1 to the single-ended port 2 and port 4, respectively. $S_{sd21}=1/\sqrt{2}(S_{21}-S_{23})$ and $S_{sd41}=1/\sqrt{2}(S_{43}-S_{41})$ are S-parameters from mixed-mode port 1 to single-ended ports 2 and 4, respectively and $f$ is the frequency.

 Similarly, the intra-pair phase skew at differential port 1 would be
\begin{align}{\label{eq_skew2}}
	t_{\rm{skew,12}} = \frac{{{phase}}(S_{sd12})}{2\pi f } - \frac{{{phase}}(S_{sd32})}{2\pi f} \,.
\end{align}

%\begin{align}{\label{eq_skew2}}
%	t_{\rm{skew,1}}=t_{1,1}-t_{2,1}\,,
%	\end{align}
%where
%\begin{align}
%	  t_{1,1}=\frac{{\rm{phase}}(S_{sd12})}{2\pi f }
% \hspace{5.5mm}\text{and}\hspace{5.5mm}  t_{2,1}=\frac{{\rm{phase}}(S_{sd32})}{2\pi f} \,.
%\end{align}

%  Correspondingly, differential insertion losses would be
%\begin{align}\label{diff_skew1}
%	S_{dd21}=\frac{1}{2}\left(|S_{21}-S_{23}|e^{i 2\pi f t_{1,2}}+|S_{43}-S_{41}|e^{i 2\pi ft_{2,2}}\right)\,,
%\end{align} 
%and
%\begin{align}\label{diff_skew2}
%	S_{dd12}=\frac{1}{2}\left(|S_{12}-S_{14}|e^{i 2\pi f t_{1,1}}+|S_{34}-S_{32}|e^{i 2\pi ft_{2,1}}\right)\,.
%\end{align} 
% 

Next, we measured the S-parameters of differential striplines in PCBs and 1.5 m DAC cables (26 AWG), which include the PCB fixture, connectors, and dual extruded twinax cable designed for 224 Gbps PAM4 applications. Using the frequency-domain definitions of intra-pair skew (See (\ref{eq_skew1})  and (\ref{eq_skew2})), we calculated skew from the measured S-parameters.
For the PCB striplines, the skew results are reciprocal, meaning they are the same left-to-right and right-to-left, $t_{\rm{skew,12}}=t_{\rm{skew,21}}$. The skew is also constant and frequency-independent, which is consistent with the parallel rising edges observed in the TDT response (Sec.~\ref{sec_TDT}).
 
 \begin{figure}[!t]
	%\centering
	\includegraphics[width=3.5in]{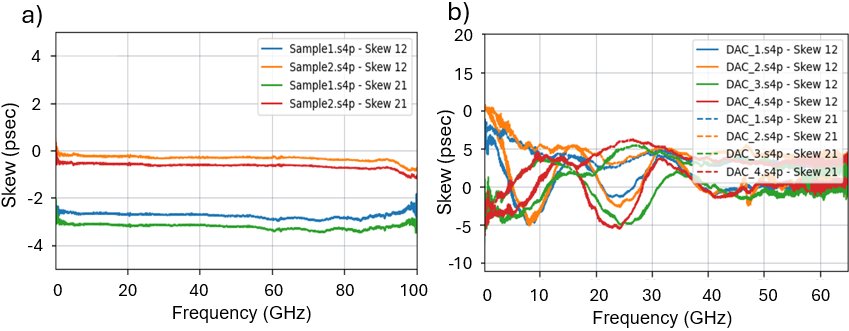}
	% where an .eps filename suffix will be assumed under latex, 
	% and a .pdf suffix will be assumed for pdflatex; or what has been declared
	% via \DeclareGraphicsExtensions.
	\caption{Intra-pair skew as a function of frequency for (a) loosely coupled striplines in PCBs and (b) strongly coupled twinax cables. }
	\label{fig_4}
	%\vspace{-1.8em}
\end{figure}

For the DAC cables, the measured skew is non-reciprocal, i.e. left-to-right and right-to-left results are different $t_{\rm{skew,12}} \neq t_{\rm{skew,21}}$. Additionally, the skew is frequency-dependent rather than constant, which aligns with the non-parallel edges observed in the TDT response (Sec.~\ref{sec_TDT}).

\begin{figure}[!t]
	%\centering
	\includegraphics[width=3.5in]{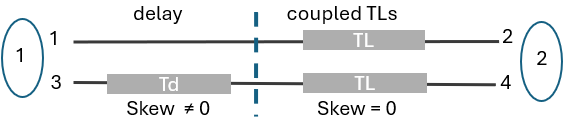}
	% where an .eps filename suffix will be assumed under latex, 
	% and a .pdf suffix will be assumed for pdflatex; or what has been declared
	% via \DeclareGraphicsExtensions.
	\caption{Schematic of Simulated Channels: intra-pair skew is introduced by placing single-ended (SE) delay lines in front of the twinax cable. 
		The delay of the delayed lines is adjusted to vary the skew in the channel from 0 to 3 ps. }
	\label{fig_5}
	%\vspace{-1.8em}
\end{figure}

\begin{figure}[!t]
	%\centering
	\includegraphics[width=3.5in]{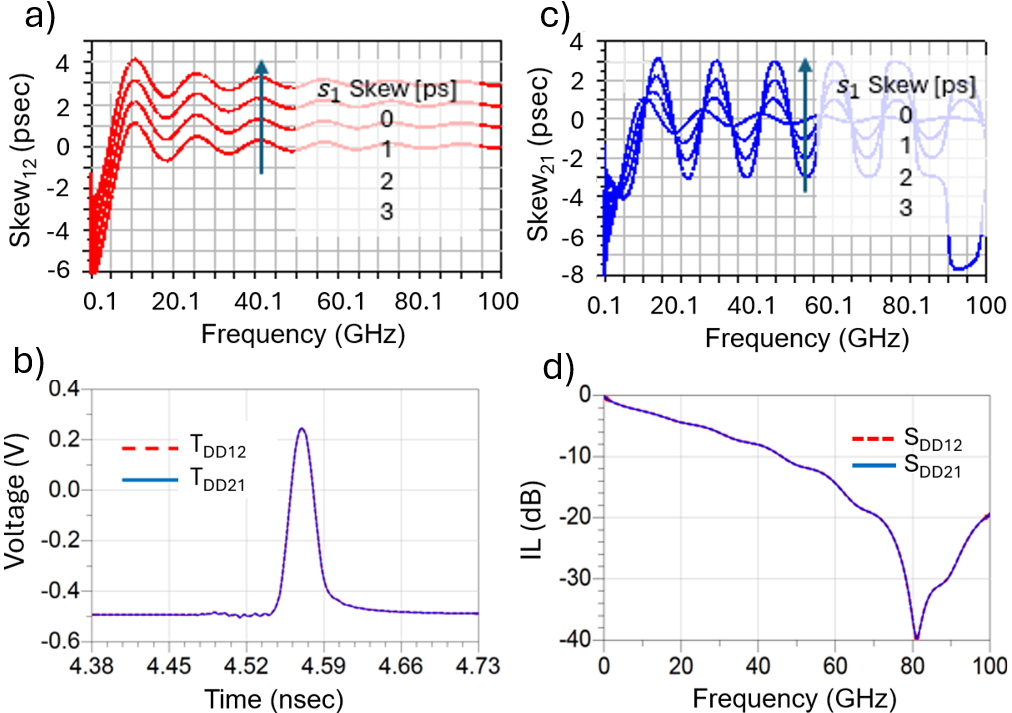}
	% where an .eps filename suffix will be assumed under latex, 
	% and a .pdf suffix will be assumed for pdflatex; or what has been declared
	% via \DeclareGraphicsExtensions.
	\caption{a) Simulated intra-pair phase skew signal propagating from right-to-left ($t_{\rm{skew,12}}$), and b) intra-pair phase skew signal propagating from left -to-right ($t_{\rm{skew,21}}$).
		Differential-to-differential responses are shown for signal propagation: left-to-right in the blue line and right-to-left in the red line c) pulse responses, and d) through insertion losses. }
	\label{fig_6}
	%\vspace{-1.8em}
\end{figure}

Next, we performed TDT and frequency-domain S-parameter simulations of a cascaded channel. The channel was created by cascading a delay line with constant skew across frequency onto the dual-extruded twinax cable (Fig.~\ref{fig_5}). The delay line was swept from 0 ps to 3 ps in 1 ps steps.
Port 1 (Fig.~\ref{fig_6}a): The skew profile shifts upward in parallel as the delay increases. Since the skew is added after the coupled transmission line, the coupling does not alter the added skew—whatever small skew the cable had is simply offset by the delay line value.
Port 2 (Fig.~\ref{fig_6}c): The behavior is very different. Here, the delayed signal passes through the coupled transmission line, which impacts the skew. The skew profile becomes oscillatory around zero, with a magnitude of the delay line skew. This occurs because of the out-of-phase oscillation between single-ended transmission and forward coupling: sometimes the signal is fully on one line, and other times it is fully on the other line.

As one can see, frequency domain intra-pair skew is not reciprocal—it is different from left-to-right versus right-to-left ($t_{\rm{skew,12}} \neq t_{\rm{skew,21}}$). However, the differential pulse response (Fig.~\ref{fig_6}b) and differential insertion loss (Fig.~\ref{fig_6}d) remain the same in both directions, showing that the overall channel response is reciprocal. Thus, while the skew defined in (\ref{eq_skew1}) and (\ref{eq_skew2}) is a useful mathematical construct, it is not a reciprocal parameter and therefore, cannot fully capture the true impact of skew in the channel.

% \begin{figure}[!t]
%	%\centering
%	\includegraphics[width=3.5in]{fig5_1.PNG}
%	% where an .eps filename suffix will be assumed under latex, 
%	% and a .pdf suffix will be assumed for pdflatex; or what has been declared
%	% via \DeclareGraphicsExtensions.
%	\caption{Schematic of a cascaded, lossless coupled transmission line (TL1) with nonzero skew and a coupled line (TL2) with zero skew. The common mode conversion flow diagram is shown for differential signal transmission from the left (a) and from the right (b).}
%	\label{fig_5_1}
%	%\vspace{-1.8em}
%\end{figure}
%
%
% \begin{figure}[!t]
%	%\centering
%	\includegraphics[width=3.5in]{fig5_2.PNG}
%	% where an .eps filename suffix will be assumed under latex, 
%	% and a .pdf suffix will be assumed for pdflatex; or what has been declared
%	% via \DeclareGraphicsExtensions.
%	\caption{Schematic of a cascaded, lossless coupled transmission line (TL1) with nonzero skew and a coupled line (TL2) with zero skew. The common mode conversion flow diagram is shown for differential signal transmission from the left (a) and from the right (b).}
%	\label{fig_5_2}
%	%\vspace{-1.8em}
%\end{figure}

 \begin{figure}[!t]
	%\centering
	\includegraphics[width=3.5in]{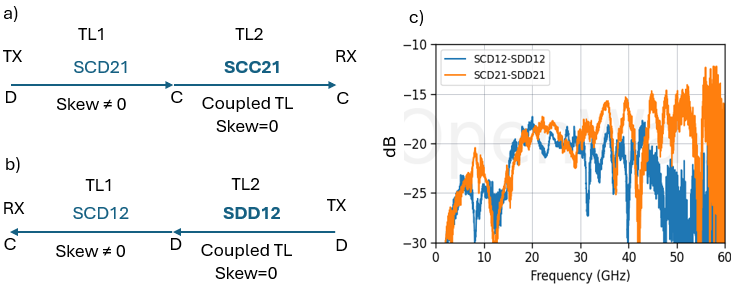}
	% where an .eps filename suffix will be assumed under latex, 
	% and a .pdf suffix will be assumed for pdflatex; or what has been declared
	% via \DeclareGraphicsExtensions.
	\caption{Schematic of a cascaded, transmission line (TL1) with nonzero skew and a coupled line (TL2) with zero skew. The common mode conversion flow diagram is shown for differential signal transmission from the left (a) and from the right (b). c) Measured
		differential to common model conversion minus differential insertion loss plotted left to right and right to left signal propagation. }
	\label{fig_5_3}
	%\vspace{-1.8em}
\end{figure}

\subsection{Differential to Common Mode Insertion Loss}

Another metric sometimes used to quantify intra-pair skew is the difference between common-mode insertion loss and differential insertion loss. Let’s examine how this metric behaves.

Consider a cascaded channel where TL1 has non-zero intra-pair skew, followed by TL2, a strongly coupled line with zero skew. We analyze the differential-to-common-mode conversion when:
A differential signal is excited at the left port and the resulting common-mode signal is received at the right port (Fig.~\ref{fig_5_3}a):
\[
%\begin{align}
	SC_2=S_{cd21}+S_{cc21}\,.
\]
%\end{align}
A differential signal is excited at the right port and the resulting common-mode signal is received at the left port (Fig.~\ref{fig_5_3}b):
\[
%\begin{align}
	SC_1=S_{cd12}+S_{dd12}\,.
\]
%\end{align}
Note that the differential to common mode conversion in delay line TL1 is assumed to be reciprocal $S_{cd21}=S_{cd12}$.  $S_{dd12}$ and $S_{cc21}$ is differential and common mode insertion losses of TL2. In the first case, the signal is attenuated by the common-mode insertion loss.
In the second case, the signal is attenuated by the differential insertion loss.
Thus, when the differential and common-mode insertion losses are not equal, the differential-to-common-mode conversion differs depending on the excitation direction. This leads to non-reciprocal behavior, which is typical in strongly coupled transmission lines, such as twinax cables.

The difference between common-mode and differential insertion losses is therefore not always reciprocal and may not be a reliable metric for quantifying the impact of skew.
For example, in Fig.~\ref{fig_5_3}c, the measured differential-to-common-mode conversions (normalized by insertion loss) from both ends of a DAC cable are shown. The results are not identical, highlighting the limitations of this metric.

 \begin{figure}[!t]
	%\centering
	\includegraphics[width=3.5in]{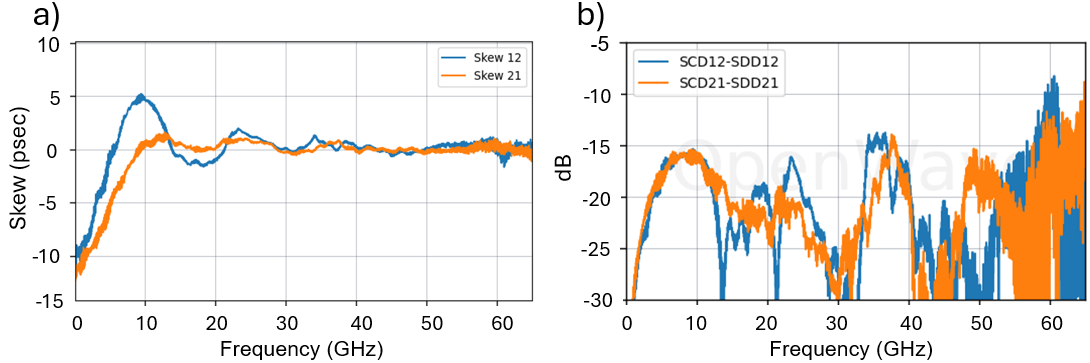}
	% where an .eps filename suffix will be assumed under latex, 
	% and a .pdf suffix will be assumed for pdflatex; or what has been declared
	% via \DeclareGraphicsExtensions.
	\caption{a) Intra-pair skews for left-to-right and right-to-left signal propagation in a 1.5 m DAC cable (26 AWG). b) Differential-to-common mode conversion minus differential insertion loss, plotted for left-to-right and right-to-left signal propagation in the same cable. }
	\label{fig_8}
	%\vspace{-1.8em}
\end{figure}

\subsection{Effective Intra-Pair Skew}
In \cite{2021_Moon_spi} authors introduce an Effective Intra-Pair Skew (EIPS) metric, calculated as a weighted average of the magnitude of skew, defined by (\ref{eq_skew1}) and (\ref{eq_skew2}):
\[
%\begin{align}
	EIPS=\int_{fmin}^{fmax}W(f)|skew(f)|df\,.
\]
%\end{align}
The weighting factors ($W$) account for de-skewing and mode conversion. Because this approach depends on non-reciprocal skew and mode conversions, EIPS itself is non-reciprocal.
Fig.~\ref{fig_8} presents an example of a measured 1.5 m DAC cable used in a 224 Gbps PAM4 application. Fig.~\ref{fig_8}a shows a skew profile that is non-symmetric, Fig.~\ref{fig_8}b shows asymmetric mode conversion and different differential insertion loss.
The EIPS values for directions left-to-right and right-to-left are 3.8 ps and 1.8 ps, respectively. The lack of symmetry underscores EIPS’s limitation as a skew-impact quantification metric.

\section{Skew Induced Insertion Loss Deviation}
As we saw in previsions section, time-domain methods are unreliable for coupled interconnects because skew differs across voltage levels and frequency domain conventional methods have issue with reciprocity/symmetry. It is problematic if a channel passes the skew specification from one 
direction but fails  from the other, since skew impact is inherently reciprocal (see Section \ref{sec_FDS}).

It is preferable to assess skew impact using a reciprocal parameter to which the SerDes 
receiver is sensitive; differential insertion loss is such a parameter.

 \begin{figure}[!t]
	%\centering
	\includegraphics[width=3.5in]{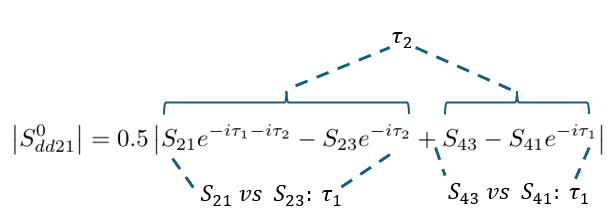}
	% where an .eps filename suffix will be assumed under latex, 
	% and a .pdf suffix will be assumed for pdflatex; or what has been declared
	% via \DeclareGraphicsExtensions.
	\caption{Schematic of S-parameters de-skewing.}
	\label{fig_9}
	%\vspace{-1.8em}
\end{figure}

\subsection{De-skewing S-Parameters}\label{sec_deskew}
Before we define skew induced insertion loss parameter we need to de-skew S-parameters. Let $\tau_1$ be 
phase delay difference between THRU and FEXT (e.g. $S_{21}$ and $S_{23}$), and $\tau_2$ between differential and single ended S-parameters (e.g. $S_{sd21}$ and $S_{sd41}$) (see Fig.~\ref{fig_9}). Then, we can de-skew S-parameters as follows
\begin{equation}
	\begin{array}{c}
		S^{0}_{21} = S_{21}e^{-i(\tau_1 + \tau_2)}, \quad S^{0}_{23} = S_{23}e^{-i\tau_2}, \\ [0.3cm]
		S^{0}_{32} = S_{32}e^{-i\tau_2}, \quad  S^{0}_{12} = S_{12}e^{-i(\tau_1 + \tau_2)}, \\ [0.3cm]
		S^{0}_{14} = S_{14}e^{-i\tau_1}, \quad  S^{0}_{41} = S_{41}e^{-i\tau_1}.
	\end{array}\label{new_s_par}
\end{equation}

Based on the above notations ~(\ref{new_s_par}) and definition of the skew likewise in (\ref{eq_skew1}) and (\ref{eq_skew2}), the following equations can be obtained for the updated skews
\begin{equation}\label{new_skew}
 \begin{array}{c}
	\displaystyle t^{0}_{skew,21} = \frac{1}{2\pi f} \left[phase\left(\left| S_{21}e^{-i\tau_1} - S_{23} \right|e^{-i(\tau_2 + \tau_3)}\right) - \right.\\ [0.2cm]
		                            \left. - phase\left(\left| S_{43} - S_{41}e^{-i\tau_1} \right|e^{-i\tau_4} \right) \right], \\  [0.3cm]
		                             
	\displaystyle t^{0}_{skew,12} = \frac{1}{2\pi f} \left[phase\left(\left| S_{12}e^{-i\tau_2} - S_{14} \right|e^{-i(\tau_1 + \tau_5)}\right) - \right.\\ [0.2cm]
                                    \left. - phase\left(\left| S_{34} - S_{14}e^{-i\tau_2} \right|e^{-i\tau_6} \right) \right],  
 \end{array}
\end{equation}
where
\[
\begin{array}{c}
	\tau_3 = phase\left(S_{21}e^{-i\tau_1} - S_{23}\right), \\ [0.15cm]
	\tau_4 = phase\left(S_{43} - S_{41}e^{-i\tau_1}\right), \\  [0.15cm]
	\tau_5 = phase\left(S_{12}e^{-i\tau_2} - S_{14}\right), \\  [0.15cm]
	\tau_6 = phase\left(S_{34} - S_{32}e^{-i\tau_2}\right).
\end{array}
\]

%\begin{equation}\label{new_skew}
%	\begin{array}{c}
%		t^{skew=0}_{skew,2} = \frac{-i}{2\pi f} \left(\tau_2 + \tau_3 - \tau_4 \right), \\  t^0_{skew,1} = \frac{-i}{2\pi f} \left(\tau_1 + \tau_5 - \tau_6\right),
%	\end{array}
%\end{equation}

Our goal is to find $\tau_1$ and $\tau_2$ such a way that the updated skews in (\ref{new_skew}) are zeros $t^{0}_{skew,21} = t^{0}_{skew,12} = 0$. Thus, we get the following system of nonlinear equations
\[
\begin{cases}
	\displaystyle	\tau_2 \!+\! phase\left(S_{21}e^{-i\tau_1} \!-\! S_{23}\right) \!-\! phase\left(S_{43} \!-\! S_{41}e^{-i\tau_1}\right) \!=\! 0 , \\
	\displaystyle	\tau_1 \!+\! phase\left(S_{12}e^{-i\tau_2} \!-\! S_{14}\right) \!-\! phase\left(S_{34} \!-\! S_{32}e^{-i\tau_2}\right) \!=\! 0 ,
\end{cases}
\]
which in turn resulting in
\begin{equation}
	\begin{cases}
		\displaystyle \tau_2 + tan^{-1}\left(\frac{\frac{Im(z_1)}{Re(z_1)}-\frac{Im(z_2)}{Re(z_2)}}{1 + \frac{Im(z_1)}{Re(z_1)}\frac{Im(z_2)}{Re(z_2)}}\right) = 0 \, , \\[0.5cm]
		\displaystyle \tau_1 + tan^{-1}\left(\frac{\frac{Im(z_3)}{Re(z_3)}-\frac{Im(z_4)}{Re(z_4)}}{1 + \frac{Im(z_3)}{Re(z_3)}\frac{Im(z_4)}{Re(z_4)}}\right) = 0 \, ,
	\end{cases}
\label{system}
\end{equation}
where
\[
\begin{array}{c}
	z_1 = S_{21}e^{-i\tau_1} - S_{23}, \quad
	z_2 = S_{43} - S_{41}e^{-i\tau_1}, \\  [0.15cm]
	z_3 = S_{12}e^{-i\tau_2} - S_{14}, \quad
	z_4 = S_{34} - S_{32}e^{-i\tau_2}.
\end{array}
\]

The phase differences $\left(\tau_1, \tau_2\right)$ now, can be determined by applying numerical methods to solve the nonlinear system (\ref{system}).

One should note that the magnitudes of the de-skewed left-to-right and right-to-left differential insertion losses will automatically be equal.
\[
\left|S^0_{dd21}\right| \!=\! 0.5\left|S_{21}e^{-i\tau_1 - i\tau_2} \!-\! S_{23}e^{-i\tau_2} \!+\! S_{43} \!-\! S_{41}e^{-i\tau_1}\right| \!=
\]
\[
=\! 0.5\left|S_{12}e^{-i\tau_1 - i\tau_2} \!-\! S_{14}e^{-i\tau_1} \!+\! S_{34} \!-\! S_{32}e^{-i\tau_2}\right| \!=\! \left|S^0_{dd12}\right|.
\]

\begin{figure}[t]
	%\centering
	\includegraphics[width=3.5in]{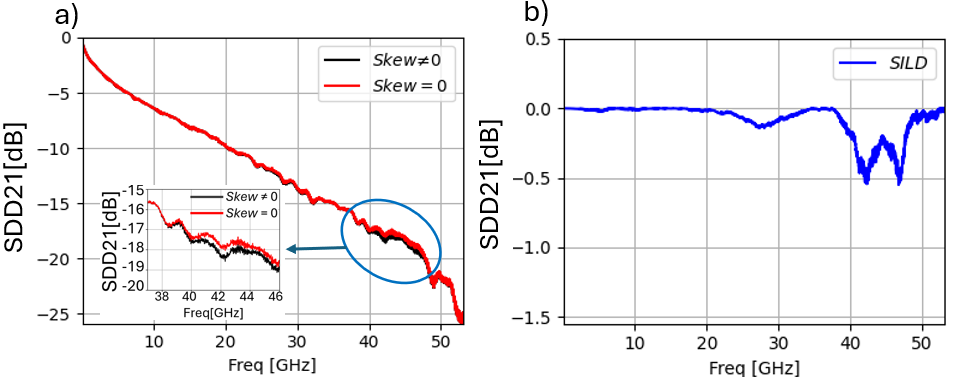}
	% where an .eps filename suffix will be assumed under latex, 
	% and a .pdf suffix will be assumed for pdflatex; or what has been declared
	% via \DeclareGraphicsExtensions.
	\caption{a) Differential insertion loss as a function of frequency before and after de-skewing.
		b) Corresponding SILD as a function of frequency.}
	\label{fig_9_1}
	%\vspace{-1.4em}
\end{figure}

\begin{figure}[t]
	%\centering
	\includegraphics[width=3.5in]{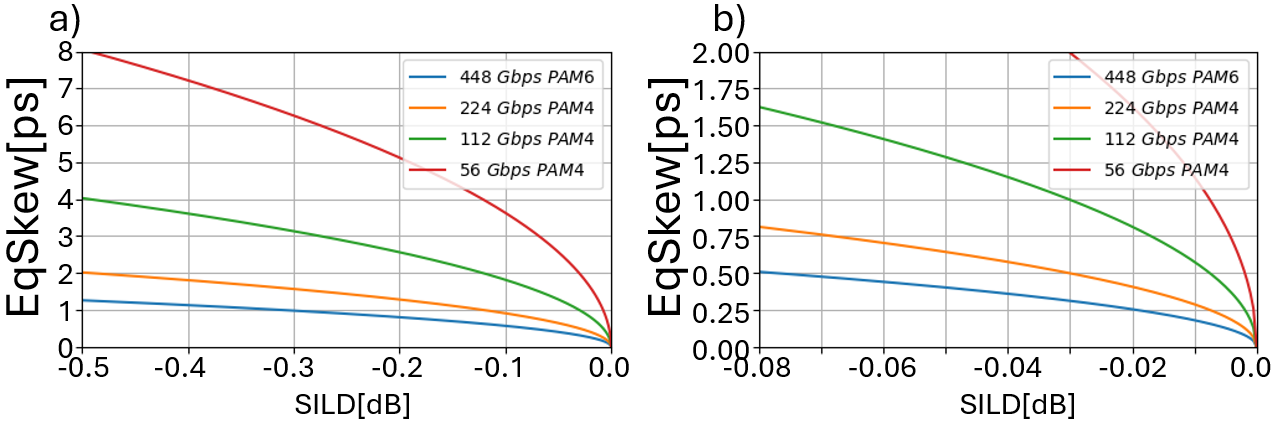}
	% where an .eps filename suffix will be assumed under latex, 
	% and a .pdf suffix will be assumed for pdflatex; or what has been declared
	% via \DeclareGraphicsExtensions.
	\caption{a) Equivalent intra-pair skew as a function of SILD for various data rates (Nyquist frequencies).
		b) Zoomed-in view of (a).}
	\label{fig_eqSkew}
	%\vspace{-1.4em}
\end{figure}

\subsection{Skew-Induced Insertion Loss Deviation as an Intra-pair skew metric} 
In this subsection, we will introduce intra-pair skew metrics. Since we can de-skew the differential insertion loss magnitude (Sec.~\ref{sec_deskew}) for either left-to-right or right-to-left transmission, we can calculate the so-called Skew-Induced Insertion Loss Deviation defined as follows
\begin{align}\label{SILD1_2_eq}
	\mathrm{SILD}_{1(2)} = |\mathrm{S}_{dd12(21)}| - |\mathrm{S}^0_{dd12(21)}|\,.
\end{align} 
SILD is reciprocal and, therefore, is the same when calculated from left-to-right or right-to-left transmission. It measures the distortion of the differential insertion loss magnitude caused by intra-pair skew. The maximum absolute value of SILD within the signal bandwidth can be used as a metric. Additionally, we can derive another single-number skew metric by calculating the RMS value of SILD, similar to the figure of merit for insertion loss deviation defined in the IEEE 802.3 standard. Thus, the figure of merit for SILD will be:
\begin{align}\label{fom_sild}
	\mathrm{FOM\_SILD_{1(2)}}=\sqrt{\frac{1}{N}\sum_{i=1}^{N}\left({W}_i *\rm{SILD_{{\it i},1(2)}}^2\right)}\,,
\end{align}
where 
\[
%\begin{align}\label{PWF}
	W_i=sinc^2\left(\frac{f_i}{f_b}\right) \frac{1}{1+\left(f_i/f_r\right)^8}\frac{1}{1+\left(f_i/f_{t}\right)^4}\,.
\]
%\end{align}
The weight function $W_i$ is defined in the IEEE 802.3 standard. Here, $f_b$ represents the signal rate, while $f_r$ and $f_{\rm{t}}$ denote the receiver 3 dB bandwidth and the 3 dB transmit filter bandwidth, respectively. The summation extends up to the maximum frequency for a given signal rate as defined in the IEEE 802.3 standard. 
$\rm{FOM\_SILD_{1(2)}}$ is reciprocal, unique number per channel and measures how much insertion loss gets distorted by the intra-pair skew.

To demonstrate the effectiveness of our de-skewing method, we present an example of measured S-parameters from a channel designed for 224 Gbps PAM4 operation. Fig.~\ref{fig_9_1}a shows both the original differential insertion loss and the de-skewed differential insertion loss. The de-skewing has the most impact around the 25 GHz and 42 GHz frequency ranges. The corresponding SILD is shown in Fig.~\ref{fig_9_1}b, with $\rm{FOM\_SILD} = 0.1 \mathrm{dB}$.

SILD defined by (\ref{SILD1_2_eq}) is the additional insertion loss caused by skew. By equating it to the loss due to skew defined by (\ref{skewloss}), one can determine the equivalent skew corresponding to a given SILD as a function of frequency:
\begin{align}\label{EqSkew}
\mathrm{EqSkew}=\frac{cos^{-1}\left(10^{\mathrm{SILD}/20}\right)}{\pi f} 1e12~\text{[ps]}\,,
\end{align}
where $f$ is the frequency in Hz.

This relationship can be used to obtain a rough estimate of the skew level corresponding to a given SILD. Figure~\ref{fig_eqSkew} shows the calculated $\mathrm{EqSkew}$ using (\ref{EqSkew}) for various Nyquist frequencies corresponding to different data rates.

\section{224 Gbps Serdes IP measured BERs versus \text{FOM\_SILDs}}

\begin{figure}[t]
	%\centering
	\includegraphics[width=3.5in]{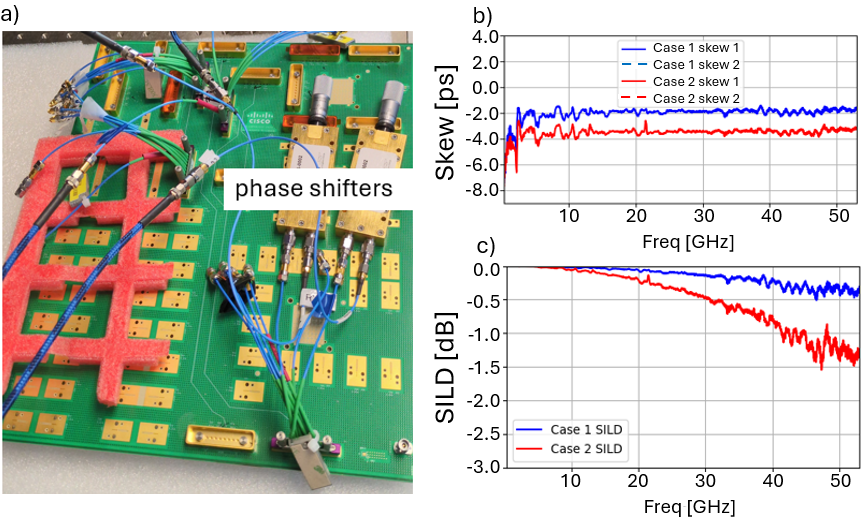}
	% where an .eps filename suffix will be assumed under latex, 
	% and a .pdf suffix will be assumed for pdflatex; or what has been declared
	% via \DeclareGraphicsExtensions.
	\caption{a) Measurement setup (DUT A) to create frequency-independent skew in the channel using phase shifters. b) Measured skew as a function of frequency for two skew values (Case 1 and Case 2) for left-to-right (skew 1) and right-to-left signal propagation (skew 2). c) SILD as a function of frequency for the two cases.  Note that, due to the symmetry of SILD from left-to-right and right-to-left, only one side is shown in the plot.}
	\label{fig_10}
	%\vspace{-1.4em}
\end{figure}

This section describes the BER measurements performed on a 224 Gbps SerDes IP. To evaluate the impact of intra-pair skew on BER, two different test channels were considered. In the first setup, Design Under Test (DUT) A was configured as follows: one end of the PCB differential stripline was connected to the inputs of two high-precision mechanical phase shifters through matched-length coaxial cables attached to the P and N traces; the other end of the stripline was connected to matched-length coaxial cables, defining one end of DUT A, while the outputs of the phase shifters were similarly connected through matched-length coaxial cables, defining the other end of DUT A. These phase shifters allowed the manual introduction of precise, frequency-independent time delays (skew) in 1 ps increments (see Fig.~\ref{fig_10}).

\begin{figure}[t]
	%\centering
	\includegraphics[width=3.5in]{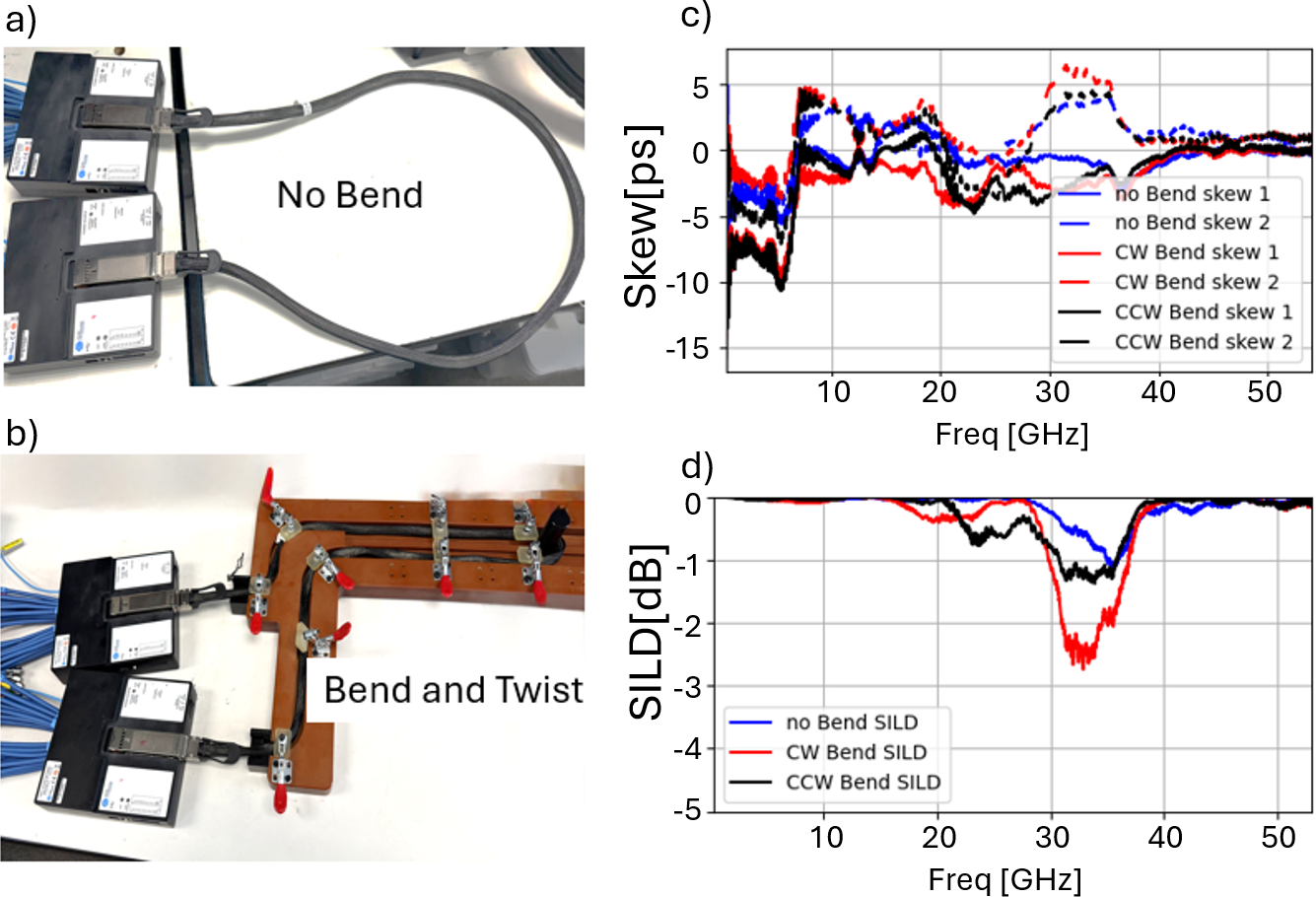}
	% where an .eps filename suffix will be assumed under latex, 
	% and a .pdf suffix will be assumed for pdflatex; or what has been declared
	% via \DeclareGraphicsExtensions.
	\caption{ a) Measurement setup for a 1.5 m DAC cable (DUT B) with no bend and no twist. b) Measurement setup for (DUT B) with bend and twist.  c) Measured skew as a function of frequency for the following cases: no bend/no twist (no bend means normal use of cable looping back to the  SerDes receiver), 360° clockwise twist (CW), 360° counterclockwise twist (CCW), and bend. d) SILD as a function of frequency for the two cases described in (c). Note that, due to the symmetry of SILD from left-to-right and right-to-left, only one side is shown in the plot.}
	\label{fig_11}
	%\vspace{-1.4em}
\end{figure}

In the second setup, DUT B consisted of a 1.5 m DAC cable assembly comprising 16 twinax cables that exhibit frequency-dependent skew characteristics and were connected to a Wilder Module Compliance Boards (MCB). The ends of the MCBs were connected to coaxial cables, which defined the terminals of DUT B (see Fig.~\ref{fig_11}).

The S-parameters of both DUTs were measured using a vector network analyzer calibrated up to 70 GHz. The initial measurement plane was established at the coaxial VNA cable interfaces using a standard Short-Open-Load-Through (SOLT) calibration. Measurements were conducted in a temperature-controlled environment at 25°C ± 2°C. Both DUTs were then connected between the SerDes transmitter and receiver, where the internal error checker was used to calculate the BER.

\begin{figure}[t]
	%\centering
	\includegraphics[width=3.5in]{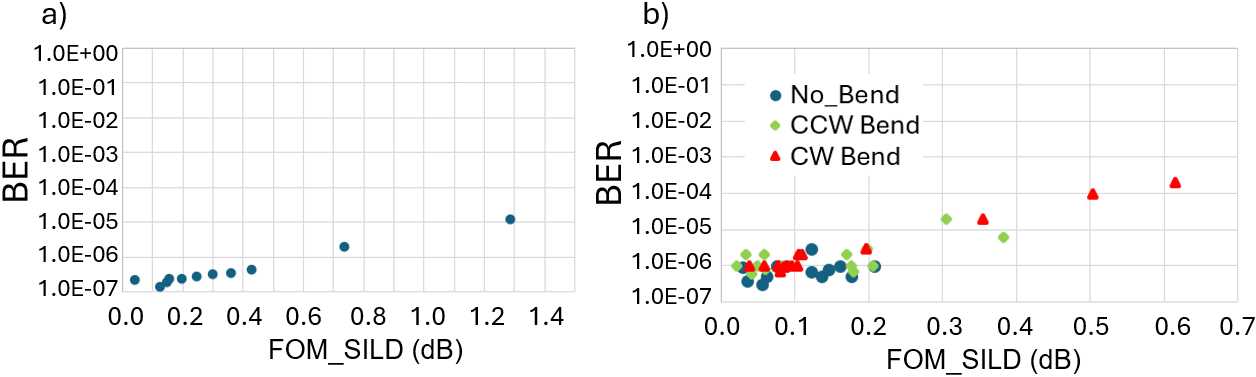}
	% where an .eps filename suffix will be assumed under latex, 
	% and a .pdf suffix will be assumed for pdflatex; or what has been declared
	% via \DeclareGraphicsExtensions.
	\caption{a) Measured BER  using a 224 Gbps SerDes IP vs $\rm{FOM\_SILD}$ for various frequency-independent intra-pair skews created in the channel using phase shifters.
		b) Measured BER using a 224 Gbps SerDes IP vs $\rm{FOM\_SILD}$ for various frequency-dependent intra-pair skews in twinax cables, including cases with no bend/no twist, and with bends incorporating 360° clockwise (CW) and 360° counterclockwise (CCW) twists.}
	\label{fig_12}
	%\vspace{-1.4em}
\end{figure}

\begin{figure}[t]
	%\centering
	\includegraphics[width=3.5in]{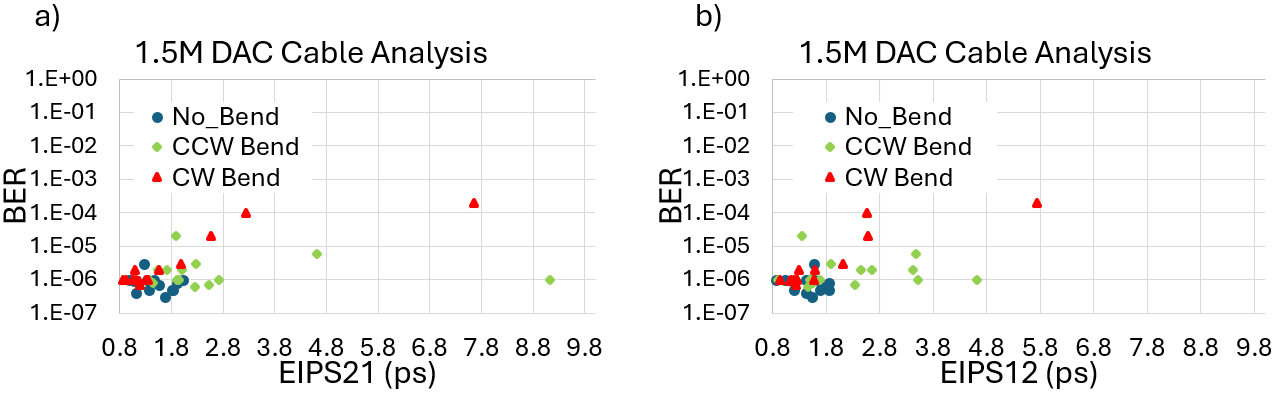}
	% where an .eps filename suffix will be assumed under latex, 
	% and a .pdf suffix will be assumed for pdflatex; or what has been declared
	% via \DeclareGraphicsExtensions.
	\caption{Measured BER using a 224 Gbps SerDes IP vs to left-to-right and right-to-left EIPS values.}
	\label{eipses}
	%\vspace{-1.4em}
\end{figure}

\begin{figure}[t]
	%\centering
	\includegraphics[width=3.5in]{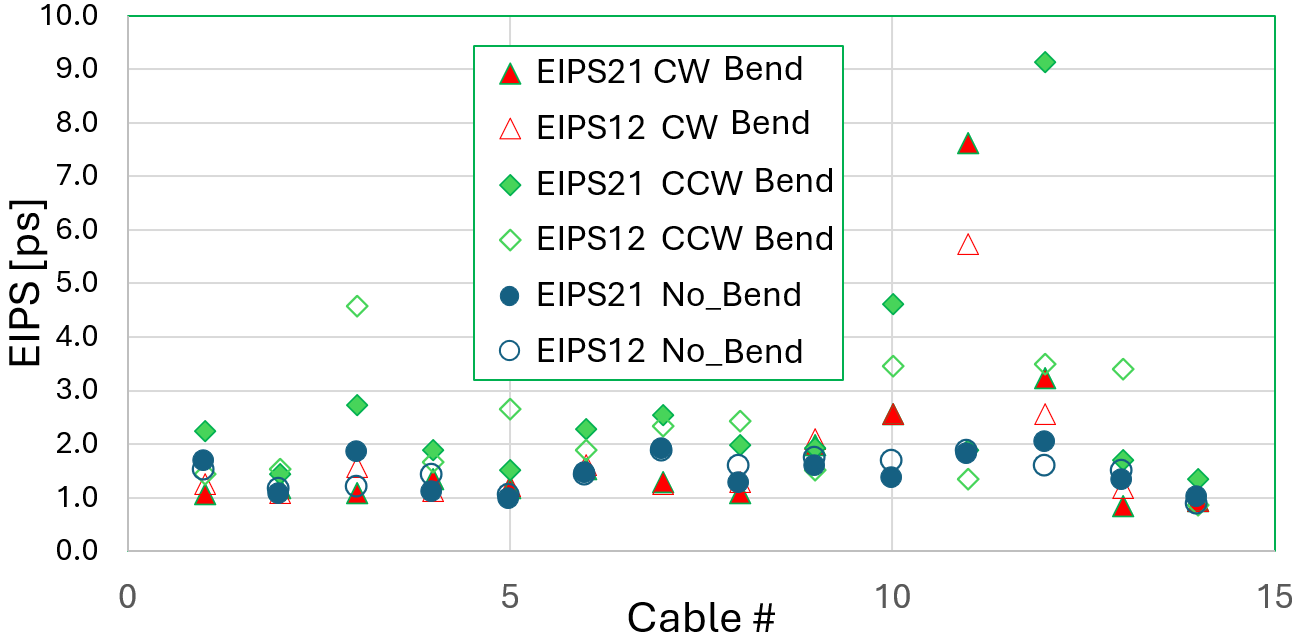}
	% where an .eps filename suffix will be assumed under latex, 
	% and a .pdf suffix will be assumed for pdflatex; or what has been declared
	% via \DeclareGraphicsExtensions.
	\caption{Correlation between left-to-right and right-to-left EIPS values.}
	\label{eips}
	%\vspace{-1.4em}
\end{figure}

\subsection{Frequency independent skew} 

In this subsection, we present BER measurements conducted using a 224 Gbps SerDes IP connected to DUT A, with the channel intra-pair skew systematically varied using phase shifters to introduce frequency-independent skew (Fig.~\ref{fig_10}). The skew was set to several distinct values, and the corresponding S-parameters were measured to calculate SILD and $\rm{FOM\_SILD}$ using (\ref{SILD1_2_eq}) and (\ref{fom_sild}), respectively. Figure~\ref{fig_10}b shows the skew calculated from the measured S-parameters using (\ref{eq_skew1}) and (\ref{eq_skew2}) as a function of frequency for two skew values (Case 1 and Case 2), for left-to-right propagation (skew 1) and right-to-left propagation (skew 2). Note that, because the skew in this case is reciprocal and identical from both sides of DUT A, the corresponding curves overlap; therefore, only two plots are visible instead of four, one for each case. The corresponding SILD responses as a function of frequency for the two cases are shown in Fig.~\ref{fig_10}c.

The measured BER as a function of $\rm{FOM\_SILD}$ is depicted in (Fig.~\ref{fig_12}a). Notably, when $\rm{FOM\_SILD}$ is below 0.3 dB, the BER remains relatively stable, indicating minimal impact on signal integrity. However, as $\rm{FOM\_SILD}$ exceeds 0.3 dB,  BER begins to increase noticeably.

In Fig. \ref{eipses}, we present the BER as a function of the left-to-right and right-to-left EIPS values under the same conditions as in Fig. \ref{fig_12}. It is evident that the EIPS values do not consistently reflect the BER performance trend. The correlation between the left-to-right and right-to-left EIPS values is shown in Fig. \ref{eips}, illustrating a substantial difference between these two measures. This discrepancy suggests that using EIPS values as a pass/fail criterion is inconclusive.

\subsection{Frequency dependent skew}

In this subsection, we present BER measurements performed using a 224 Gbps SerDes IP connected to DUT B, which exhibits frequency-dependent skew (Fig.~\ref{fig_11}). DUT B was tested under three conditions: 1) no bend/no twist, 2) 360° clockwise twist with bend, and 3) 360° counterclockwise twist with bend. The measured BER is plotted as a function of $\rm{FOM\_SILD}$ in Fig.~\ref{fig_12}b. The S-parameters of DUT B were measured to calculate SILD and $\rm{FOM\_SILD}$ using (\ref{SILD1_2_eq}) and (\ref{fom_sild}), respectively. Figure~\ref{fig_12}c shows the measured skew, calculated from the measured S-parameters using (\ref{eq_skew1}), as a function of frequency for the following cases: no bend/no twist, 360° clockwise twist (CW) with bend, and 360° counterclockwise twist (CCW) with bend. Here, "no bend" refers to the normal cable configuration, in which the cable is looped without any sharp bend or twist. Figure~\ref{fig_12}d presents SILD as a function of frequency for these cases.

The results indicate that, for $\rm{FOM\_SILD}$ values below 0.3 dB, BER remains largely unchanged, whereas a significant increase in BER is observed once $\rm{FOM\_SILD}$ exceeds 0.3 dB. Cables subjected to bending and twisting exhibit higher $\rm{FOM\_SILD}$ values and correspondingly degraded BER performance.

\section{Statistical Distribution of measured \text{FOM\_SILDs}}

To validate the practical relevance of the proposed metrics in high-volume manufacturing environments, a large-scale statistical analysis was conducted on over 3,000 high-speed differential pairs. The dataset comprises measurements from 1.5 m DAC cables, utilizing both 26 and 27 AWG twinax conductors, designed specifically for 224 Gbps PAM4 applications. These measurements are comprehensive, accounting for the entire signal path, including PCB test fixtures, I/O connectors, and the twinax cable itself.

\begin{figure}[t]
	%\centering
	\includegraphics[width=3.5in]{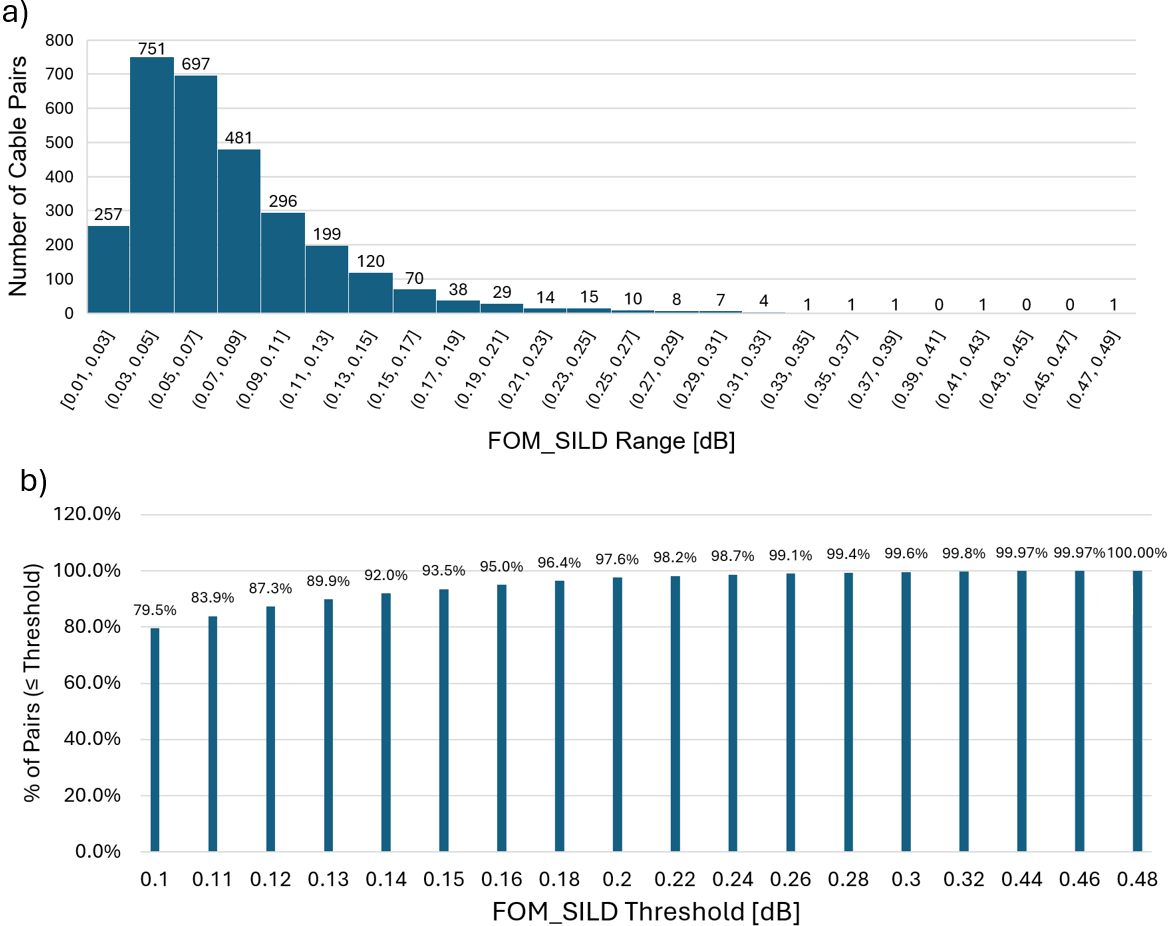}
	% where an .eps filename suffix will be assumed under latex, 
	% and a .pdf suffix will be assumed for pdflatex; or what has been declared
	% via \DeclareGraphicsExtensions.
	\caption{Measured 1.5 m 26 and 27 AWG twinax DAC cables over 3000 pairs (for 224 Gbps PAM4 applications). a) Number of cable pairs per $\rm{FOM\_SILD}$ range. b) Percentage of Cable pairs with $\rm{FOM\_SILD}$ below threshold.}
	\label{fig_13}
	%\vspace{-1.4em}
\end{figure}

The distribution of the calculated $\rm{FOM\_SILD}$ values, as illustrated in Fig.~\ref{fig_13}a, provides a clear window into the consistency of modern interconnect manufacturing. The values range from a near-ideal 0 dB to a maximum of approximately 0.49 dB. A detailed analysis of the data reveals a mean $\rm{FOM\_SILD}$ value of approximately 0.073 dB, with a standard deviation ($\sigma$) estimated at 0.002 dB. The relatively low standard deviation relative to the mean indicates that the manufacturing process for these ultra-high-speed cables is highly controlled, with the majority of pairs exhibiting minimal skew-induced distortion.

However, the "tail" of the distribution—extending from 0.2 dB to 0.49 dB—is of particular interest for signal integrity engineers. While the majority of the population (90$\%$) maintains an $\rm{FOM\_SILD}$ below 0.13 dB (Fig.~\ref{fig_13}b), the remaining 10$\%$ represents the variability inherent in complex cable assemblies. This variability is typically driven by physical asymmetries, such as the wire being off-center within the dielectric (as shown in Fig.~\ref{fig_1}b) or non-uniformities in the dual-extrusion process. In the context of 224 Gbps signaling, where the UI is extremely narrow, these outliers represent the primary risk to system-level BER performance.

By correlating these statistical findings with the SerDes BER measurements presented in Section IV, we can establish a data-driven "guard band" for channel qualification. The experimental results in Fig.~\ref{fig_12} demonstrated that BER performance remains stable when $\text{FOM}\_{\text{SILD}}$ is below the 0.2–0.3 dB range. Statistically, more than 98$\%$ of the 3,000 measured pairs fall safely below this 0.2 dB threshold. The small fraction of cables exceeding 0.3 dB correlates directly with the region where BER degradation becomes non-linear and significant.

The inclusion of standard deviation and percentile thresholds (such as the 90th and 99th percentiles) allows designers to move beyond "pass/fail" binary metrics and instead adopt a risk-based approach to channel budgeting. For instance, if an end-to-end system budget allocates 0.3 dB for total $\rm{FOM\_SILD}$, and the cable segment consumes a 99th percentile value of 0.2 dB, the remaining 0.1 dB must be strictly partitioned among the host PCB and connectors. This statistical depth ensures that $\rm{FOM\_SILD}$ serves not only as a diagnostic tool for individual failures but also as a predictive metric for yield optimization in ultra-high-speed data center interconnects.

\section{Conclusion}
This work has addressed a critical gap in signal-integrity analysis of high-speed, coupled interconnects by introducing Skew-Induced Insertion Loss Deviation (SILD) and its scalar counterpart, $\rm{FOM\_SILD}$, as fundamentally reciprocal and quantitative measures of intra-pair skew effects. Traditional time-domain techniques falter under skew and voltage-level variations, while frequency-domain methods often misalign with the bidirectional symmetry inherent to these channels. By contrast, both SILD and $\rm{FOM\_SILD}$ offer robust reciprocity, enabling consistent assessment regardless of signal direction.
We demonstrated strong correlation between  $\rm{FOM\_SILD}$  trends and BER measured using 224 Gb/s SerDes IP, showing clear and reliable correlation between $\rm{FOM\_SILD}$ and BER. Channels with $\rm{FOM\_SILD}$ below approximately 0.2–0.3 dB consistently showed stable BER performance, whereas channels exceeding this threshold exhibited sharply increasing BER. These trends validate $\rm{FOM\_SILD}$ as not only a theoretical but also a practically predictive indicator of signal-integrity degradation.
We also presented measurements of over 3,000 DAC cable channels—consisting of fixture PCBs, I/O connectors, and 1.5 m 26/27 AWG twinax cables. Notably, the majority of measured channels achieved very low $\rm{FOM\_SILD}$ values, with most falling below 0.1 dB and around 90$\%$ under 0.13 dB.

Based on these findings, one could establish channel qualification criteria by specifying: 1) a maximum allowable SILD within the signal bandwidth, and 2) a maximum acceptable value for $\rm{FOM\_SILD}$, ensuring that intra-pair skew effects remain within tolerable limits.
A global $\rm{FOM\_SILD}$ threshold, for example, around 0.2-0.3 dB, could be used as a channel qualification criterion to ensure reliable BER performance. However, if specifications are required for each segment of an end-to-end channel, the total budget (e.g. 0.3 dB) would need to be allocated across all channel segments. 

The accuracy of the calculated $\mathrm{SILD}$ and $\rm{FOM\_SILD}$ is sensitive to the reciprocity quality of the measured S-parameters. To ensure reliable results, reciprocity should be at least 99$\%$ (as defined by the IEEE P370 standard). If this level cannot be achieved, it is recommended to enforce reciprocity.

In summary, SILD and $\rm{FOM\_SILD}$ provide a compelling, reciprocal, and physically grounded framework for quantifying intra-pair skew impacts in emerging ultra-high-speed SerDes interconnects. The introduction of these metrics paves the way for improved design margins, minimizes false failures in meeting channel intra-pair skew specifications, and ensures more reliable and robust high-speed performance.

%\section*{Acknowledgments}
%This should be a simple paragraph before the References to thank those individuals and institutions who have supported your work on this article.

%{\appendix[Proof of the Zonklar Equations]
%Use $\backslash${\tt{appendix}} if you have a single appendix:
%Do not use $\backslash${\tt{section}} anymore after $\backslash${\tt{appendix}}, only $\backslash${\tt{section*}}.
%If you have multiple appendixes use $\backslash${\tt{appendices}} then use $\backslash${\tt{section}} to start each appendix.
%You must declare a $\backslash${\tt{section}} before using any $\backslash${\tt{subsection}} or using $\backslash${\tt{label}} ($\backslash${\tt{appendices}} by itself
% starts a section numbered zero.)}


\bibliographystyle{IEEEtran}
\bibliography{b}

@Article{2017_Baek_DC,
  Title                    = {New technique to quantify differential P/N glass weave skew for effective system design},
  Author                   = {S. Baek and A. Koul and K. Nalla and M. Sapozhnikov and Y. Yang and G. Maghlakelidze and J. Fan},
  Journal                  = {DesignCon},
  Year                     = {2017}
}

@Article{2017_Nozadze_epeps,
  Title                    = {Effect of time delay skew on differential insertion loss in weak and strong coupled PCB traces},
  Author                   = {D. Nozadze and A. Koul and K. Nalla and M. Sapozhnikov and V. Khilkevich},
  Journal                  = {2017 IEEE 26th Conference on Electrical Performance of Electronic Packaging and Systems (EPEPS)},
  Year                     = {2017},
}

@Article{2021_Moon_spi,
  Title                    = {Intra-Pair Skew Metric, EIPS (Effective Intra-Pair Skew)},
  Author                   = {Se-Jung Moon  and et al.},
  Journal                  = {2021 IEEE 25th Workshop on Signal and Power Integrity (SPI)},
  Year                     = {2021},
}

@Article{2007_Loyer_CT,
  Title                    = {Fiber Weave Effect: Practical Impact Analysis and Mitigation Strategies},
  Author                   = {J. Loyer and R. Kunze and X. Ye},
  Journal                  = {White paper, Circuit tree},
  Year                     = {2007}
}

@Article{2010_Miller_DC,
  Title                    = {Additional Trace Losses due to Glass- Weave Periodic Loading},
  Author                   = {J. Miller and G. Blando and I. Novak},
  Journal                  = {DesignCon},
  Year                     = {2010}
}

@Article{Nalla_2017_EMC,
  Title                    = {Measurement and correlation-based methodology for estimating worst-case skew due to glass weave effect},
  Author                   = {Kartheek Nalla and Giorgi Maghlakelidze and Amendra Koul and Seungyong Baek and Mike Sapozhnikov and Jun Fan},
  Journal                  = {EMC symposium},
  Year                     = {2017}
}

@Article{Nozadze_2017_EMC,
  Title                    = {Effective channel budget technique for high-speed channels due to differential P/N skew},
  Author                   = {D. Nozadze and A. Koul and K. Nalla and M. Sapozhnikov and V. Khilkevich},
  Journal                  = {EMC symposium},
  Year                     = {2017}
}

@Article{2018_Koul_DC,
  Title                    = {Fiber weave effect: Modeling, measurements, challenges and its impact on differential insertion loss for weak and strong-coupled differential transmission lines},
  Author                   = {A. Koul and K. Nalla and D. Nozadze and M. Sapozhnikov and Y. Yang},
  Journal                  = {DesignCon},
  Year                     = {2018}
}

@ARTICLE{2023_Liu_IEEE_TR,
  author={Liu, Yuanzhuo and Bai, Siqi and Li, Chaofeng and De Moura, Vanine Sabino and Chen, Bichen and Venkataraman, Srinivas and Wang, Xu and Kim, DongHyun},
  journal={IEEE Transactions on Signal and Power Integrity}, 
  title={Inhomogeneous Dielectric Induced Skew Modeling of Twinax Cables}, 
  year={2023},
  volume={2},
  number={},
  pages={94-102},
  doi={10.1109/TSIPI.2023.3278613}}

@Article{2014_Tian_simp,
  Title                    = {Numerical investigation of glass-weave effects on high-speed interconnects in printed circuit board},
  Author                   = {Xinxin Tian and Yao-Jiang Zhang and Jane Lim and Kelvin Qiu and Rick Brooks and Ji Zhang and Jun Fan},
  Journal                  = {2014 IEEE International Symposium on Electromagnetic Compatibility (EMC), Raleigh, NC},
  Year                     = {2014},
  Pages                    = {475-479}
}

\vfill

\end{document}